\documentclass[11pt]{article}
\usepackage{amssymb,amsmath,amsfonts}
\usepackage{graphicx}
\usepackage{graphics}
\usepackage{eepic,epsfig}

\begin{document}

\title{Electromagnetic field generated by a charge moving along a helical
orbit inside a dielectric cylinder}
\author{A. A. Saharian$^{1,2}$\thanks{%
Email address: saharyan@server.physdep.r.am}, \, A. S. Kotanjyan$^{1}$, M.
L. Grigoryan$^{1}$ \\
\textit{$^{1}$Institute of Applied Problems in Physics, 375014 Yerevan,
Armenia}\\
\textit{$^2$Departamento de F\'{\i}sica-CCEN, Universidade Federal da Para%
\'{\i}ba,} \\
\textit{58.059-970, Caixa Postal 5.008, Jo\~{a}o Pessoa, PB, Brazil}}
\date{\today}
\maketitle

\begin{abstract}
The electromagnetic field generated by a charged particle moving along a
helical orbit inside a dielectric cylinder immersed into a homogeneous
medium is investigated. Expressions are derived for the electromagnetic
potentials, electric and magnetic fields in the region inside the cylinder.
The parts corresponding to the radiation field are separated. The radiation
intensity on the lowest azimuthal mode is studied.
\end{abstract}

\bigskip

PACS number(s): 41.60.Ap, 41.60.Bq

\bigskip

\section{Introduction}

\label{sec:oscint}

The radiation from a charged particle moving along a helical orbit in vacuum
has been widely discussed in literature (see, for instance, \cite%
{Bord99,Hofm04} and references given therein). This type of electron motion
is used in helical undulators for generating electromagnetic radiation in a
narrow spectral interval at frequencies ranging from radio or millimeter
waves to X-rays. The unique characteristics, such as high intensity and high
collimation, have resulted in extensive applications of this radiation in a
wide variety of experiments and in many disciplines. These applications
motivate the importance of investigations for various mechanisms of
controlling the radiation parameters. From this point of view, it is of
interest to consider the influence of a medium on the spectral and angular
distributions of the radiation.

It is well known that the presence of medium can essentially change the
characteristics of the high energy electromagnetic processes and gives rise
to new types of phenomena. Well-known examples are Cherenkov, transition,
and diffraction radiations. In a series of papers \cite{Grig95}-\cite{Grig05}
the synchrotron radiation is considered from a charge rotating around a
dielectric ball/cylinder enclosed by a homogeneous medium. It has been shown
that the superposition between the synchrotron and Cherenkov radiations
leads to interesting effects: under the Cherenkov condition for the material
of the ball/cylinder and the particle velocity, strong narrow peaks appear
in the radiation intensity. At these peaks the radiated energy exceeds the
corresponding quantity in the case of a homogeneous medium by several orders
of magnitude. The influence of a dielectric cylinder on the radiation from a
longitudinal charged oscillator moving with constant drift velocity along
the axis of the cylinder has been considered in \cite{Saha03,Saha04}.

In a previous paper \cite{Saha05} we have investigated the
radiation by a charged particle moving along a helical orbit
inside a dielectric cylinder immersed into a homogeneous medium
(for the radiation in a dispersive homogeneous medium see
\cite{Gevo84}). Specifically, formulae are derived for the
electromagnetic fields and for the spectral-angular distribution
of the radiation intensity in the exterior medium. A special case
of the relativistic motion along the direction of the cylinder
axis with non-relativistic transverse velocity is discussed in
detail and various regimes for the undulator parameter are
considered. In the present paper we consider the electromagnetic
fields generated inside the cylinder by the charge moving along a
helical orbit. The paper is organized as follows. In the next
section, by using the Green function, the vector potential and
electromagnetic fields are determined for the region inside the
cylinder. The analytic structure of the Fourier components of the
fields are investigated in Sec. \ref{sec:radfields} and the
radiation part of the field is separated. A formula is derived for
the radiation intensity on the lowest azimuthal mode. Sec.
\ref{sec:Conc} concludes the main results of the paper. In
Appendix we derive a summation formula over the eigenmodes of the
dielectric cylinder which is used in Sec. \ref{sec:radfields} for
the evaluation of the mode sums in the formulae of the radiation
intensities.

\section{Electromagnetic potentials and fields inside a cylinder}

\label{sec:oscfields}

We consider a dielectric cylinder of radius $\rho _{1}$ and with dielectric
permittivity $\varepsilon _{0}$ and a point charge $q$ moving along the
helical trajectory of radius $\rho _{0}<\rho _{1}$. We assume that the
system is immersed in a homogeneous medium with permittivity $\varepsilon
_{1}$. The velocities of the charge along the axis of the cylinder (drift
velocity) and in the perpendicular plane we will denote by $v_{\parallel }$
and $v_{\perp }$, respectively. In a properly chosen cylindrical coordinate
system ($\rho ,\phi ,z$) with the $z$-axis along the cylinder axis, the
components of the current density created by the charge are given by the
formula
\begin{equation}
j_{l}=\frac{q}{\rho }v_{l}\delta (\rho -\rho _{0})\delta (\phi -\omega
_{0}t)\delta (z-v_{\parallel }t),  \label{hosqxtutjun}
\end{equation}%
where $\omega _{0}=v_{\perp }/\rho _{0}$ is the angular velocity of the
charge. This type of motion can be produced by a uniform constant magnetic
field directed along the axis of a cylinder, by a circularly polarized plane
wave, or by a spatially periodic transverse magnetic field of constant
absolute value and a direction that rotates as a function of the coordinate $%
z$. In the helical undulators the last configuration is used.

The vector potential of the electromagnetic field is expressed in terms of
the Green function $G_{il}(\mathbf{r},t,\mathbf{r}^{\prime },t^{\prime })$
as:
\begin{equation}
A_{i}(\mathbf{r},t)=-\frac{1}{2\pi ^{2}c}\int G_{il}(\mathbf{r},t,\mathbf{r}%
^{\prime },t^{\prime })j_{l}(\mathbf{r}^{\prime },t)d\mathbf{r}^{\prime
}dt^{\prime },  \label{vecpot}
\end{equation}%
where the summation over $l$ is understood. For static and cylindrically
symmetric medium the Green function is presented in the form of the Fourier
expansion
\begin{equation}
G_{il}(\mathbf{r},t,\mathbf{r}^{\prime },t^{\prime })=\sum_{m=-\infty
}^{\infty }\int_{-\infty }^{\infty }dk_{z}d\omega G_{il}(m,k_{z},\omega
,\rho ,\rho ^{\prime })\exp [im(\phi -\phi ^{\prime })+ik_{z}(z-z^{\prime
})-i\omega (t-t^{\prime })].  \label{GF_furieexp}
\end{equation}%
The substitution of expressions (\ref{hosqxtutjun}) and (\ref{GF_furieexp})
into formula (\ref{vecpot}) gives the following result for the components of
the vector potential:
\begin{eqnarray}
A_{l}(\mathbf{r},t) &=&-\frac{q}{\pi c}\sum_{m=-\infty }^{\infty }e^{im(\phi
-\omega _{0}t)}\int_{-\infty }^{\infty }dk_{z}e^{ik_{z}(z-v_{\parallel }t)}
\notag \\
&&\times \left[ v_{\perp }G_{l\phi }(m,k_{z},\omega _{m}(k_{z}),\rho ,\rho
_{0})+v_{\parallel }G_{lz}(m,k_{z},\omega _{m}(k_{z}),\rho ,\rho _{0})\right]
,  \label{vecpot1}
\end{eqnarray}%
where
\begin{equation}
\omega _{m}(k_{z})=m\omega _{0}+k_{z}v_{\parallel }.  \label{omegam}
\end{equation}%
In the Lorentz gauge, using the formula for the Green function given earlier
in Ref. \cite{Grig95} and introducing the notations
\begin{equation}
\lambda _{j}^{2}=\frac{\omega _{m}^{2}(k_{z})}{c^{2}}\varepsilon
_{j}-k_{z}^{2},\quad j=0,1,  \label{lambdaj}
\end{equation}%
for the corresponding Fourier components $G_{il}=G_{il}(m,k_{z},\omega
_{m}(k_{z}),\rho ,\rho _{0})$ in the region inside the cylinder, $\rho <\rho
_{1}$, we obtain
\begin{subequations}
\label{GFcomp}
\begin{eqnarray}
G_{l\phi } &=&\frac{\pi }{4i^{\sigma _{l}-1}}\sum_{p=\pm 1}p^{\sigma _{l}}%
\left[ J_{m+p}(\lambda _{0}\rho _{<})H_{m+p}(\lambda _{0}\rho
_{>})+B_{m}^{(p)}J_{m+p}(\lambda _{0}\rho )\right] ,\;l=\rho ,\phi ,
\label{GFcompa} \\
G_{lz} &=&-\frac{k_{z}}{2i^{\sigma _{l}}}\frac{J_{m}(\lambda _{0}\rho
_{0})H_{m}(\lambda _{1}\rho _{1})}{\rho _{1}\alpha _{m}W_{m}^{J}}\sum_{p=\pm
1}\frac{J_{m+p}(\lambda _{0}\rho )}{p^{\sigma _{l}-1}W_{m+p}^{J}}%
H_{m+p}(\lambda _{1}\rho _{1}),\;l=\rho ,\phi ,  \label{GFcompb} \\
G_{zz} &=&\frac{\pi }{2i}\left[ J_{m}(\lambda _{0}\rho _{<})H_{m}(\lambda
_{0}\rho _{>})-J_{m}(\lambda _{0}\rho _{0})J_{m}(\lambda _{0}\rho )\frac{%
W_{m}^{H}}{W_{m}^{J}}\right] ,  \label{GFcompc}
\end{eqnarray}%
where $J_{m}(x)$ is the Bessel function and $H_{m}(x)=H_{m}^{(1)}(x)$ is the
Hankel function of the first kind, $\rho _{<}=\min (\rho ,\rho _{0})$, $\rho
_{>}=\max (\rho ,\rho _{0})$, and $\sigma _{\rho }=1$, $\sigma _{\phi }=2$.
The coefficients $B_{m}^{(p)}$ in these formulae are determined by the
expressions
\end{subequations}
\begin{equation}
B_{m}^{(p)}=-J_{m+p}(\lambda _{0}\rho _{0})\frac{W_{m+p}^{H}}{W_{m+p}^{J}}+%
\frac{ip\lambda _{1}H_{m+p}(\lambda _{1}\rho _{1})}{\pi \rho _{1}\alpha
_{m}W_{m+p}^{J}}H_{m}(\lambda _{1}\rho _{1})\sum_{l=\pm 1}\frac{%
J_{m+l}(\lambda _{0}\rho _{0})}{W_{m+l}^{J}},  \label{Bm+p}
\end{equation}%
where
\begin{equation}
\alpha _{m}=\frac{\varepsilon _{0}}{\varepsilon _{1}-\varepsilon _{0}}-\frac{%
1}{2}\lambda _{0}J_{m}(\lambda _{0}\rho _{1})\sum_{l=\pm 1}l\frac{%
H_{m+l}(\lambda _{1}\rho _{1})}{W_{m+l}^{J}}.  \label{bet1}
\end{equation}%
Here and below we use the notation%
\begin{equation}
W_{m}^{F}=F_{m}(\lambda _{0}\rho _{1})\frac{\partial H_{m}(\lambda _{1}\rho
_{1})}{\partial \rho _{1}}-H_{m}(\lambda _{1}\rho _{1})\frac{\partial
F_{m}(\lambda _{0}\rho _{1})}{\partial \rho _{1}}.  \label{WronskianF}
\end{equation}%
with $F=J,H$. Taking $\varepsilon _{1}=\varepsilon _{0}$, from (\ref{GFcomp}%
) we obtain the Fourier components of the Green function in a homogeneous
medium. In this limit $B_{m}^{(p)}=0$ and $W_{m}^{H}=0$. By using formulae (%
\ref{GFcomp}), the electromagnetic fields can be evaluated for an arbitrary
motion of a charged particle with zero radial velocity.

In the case of the motion along helical orbit, substituting the expressions
for the components of the Green function into formula (\ref{vecpot1}), we
present the vector potential in the form of the Fourier expansion
\begin{equation}
A_{l}(\mathbf{r},t)=\sum_{m=-\infty }^{\infty }e^{im(\phi -\omega
_{0}t)}\int_{-\infty }^{\infty }dk_{z}e^{ik_{z}(z-v_{\parallel
}t)}A_{ml}(k_{z},\rho ).  \label{vecpot2}
\end{equation}%
As the functions $A_{l}(\mathbf{r},t)$\ are real one has $A_{ml}^{\ast
}(k_{z},\rho )=A_{-ml}(-k_{z},\rho )$. Consequently, formula (\ref{vecpot2})
can also be rewritten in the form
\begin{equation}
A_{l}(\mathbf{r},t)=2{\mathrm{Re}}\sideset{}{'}{\sum}_{m=0}^{\infty
}e^{im(\phi -\omega _{0}t)}\int_{-\infty }^{\infty
}dk_{z}e^{ik_{z}(z-v_{\parallel }t)}A_{ml}(k_{z},\rho ),  \label{vecpot21}
\end{equation}%
where the prime means that the term $m=0$ should be taken with the weight
1/2. In the discussion below we will assume that $m\geqslant 0$. The Fourier
components $A_{ml}=A_{ml}(k_{z},\rho )$ can be presented in the form
\begin{equation}
A_{ml}=A_{ml}^{(0)}+A_{ml}^{(1)},  \label{Adecompose}
\end{equation}%
where $A_{ml}^{(0)}$ corresponds to the field of the charge in a homogeneous
medium with permittivity $\varepsilon _{0}$. For this part one has
\begin{subequations}
\label{vecpot30}
\begin{eqnarray}
A_{ml}^{(0)} &=&-\frac{qv_{\perp }}{4ci^{\sigma _{l}-1}}\sum_{p=\pm
1}p^{\sigma _{l}}J_{m+p}(\lambda _{0}\rho _{<})H_{m+p}(\lambda _{0}\rho
_{>}),\;l=\rho ,\phi ,  \label{vecpot30a} \\
A_{mz}^{(0)} &=&\frac{qv_{\parallel }i}{2c}J_{m}(\lambda _{0}\rho
_{<})H_{m}(\lambda _{0}\rho _{>}).  \label{vecpot30b}
\end{eqnarray}%
The part due to the presence of the cylinder is determined by the formulae
\end{subequations}
\begin{subequations}
\label{vecpot3}
\begin{eqnarray}
A_{ml}^{(1)} &=&\frac{q}{2\pi i^{\sigma _{l}}}\sum_{p=\pm 1}p^{\sigma
_{l}}C_{m}^{(p)}J_{m+p}(\lambda _{0}\rho ),\;l=\rho ,\phi ,  \label{vecpot3a}
\\
A_{mz}^{(1)} &=&-\frac{iqv_{\parallel }}{2c}\frac{W_{m}^{H}}{W_{m}^{J}}%
J_{m}(\lambda _{0}\rho _{0})J_{m}(\lambda _{0}\rho ),  \label{vecpot3b}
\end{eqnarray}%
with the coefficients
\end{subequations}
\begin{equation}
C_{m}^{(p)}=-\frac{i\pi v_{\perp }}{2c}B_{m}^{(p)}+pv_{\parallel }k_{z}\frac{%
J_{m}(\lambda _{0}\rho _{0})H_{m+p}(\lambda _{1}\rho _{1})H_{m}(\lambda
_{1}\rho _{1})}{c\rho {_{1}}\alpha _{m}W_{m}^{J}W_{m+p}^{J}}.  \label{Cm+p}
\end{equation}%
The electric and magnetic fields are obtained by means of standard formulae
of electrodynamics. As is seen from formula (\ref{vecpot2}), analogous
expressions may also be written for these fields. For the Fourier components
of the magnetic field one has%
\begin{equation}
H_{ml}(k_{z},\rho )=H_{ml}^{(0)}+H_{ml}^{(1)},  \label{Hdecompose}
\end{equation}%
where $H_{ml}^{(0)}$ corresponds to the field of the charge in a homogeneous
medium with permittivity $\varepsilon _{0}$. For $\rho >\rho _{0}$ this part
is determined by the formulae
\begin{subequations}
\label{magnetic0}
\begin{eqnarray}
H_{ml}^{(0)} &=&-\frac{qk_{z}}{2\pi i^{\sigma _{l}}}\sum_{p=\pm 1}p^{\sigma
_{l}-1}D_{m}^{(0p)}H_{m+p}(\lambda _{0}\rho ),\;l=\rho ,\phi ,
\label{magnetic0a} \\
H_{mz}^{(0)} &=&-\frac{q\lambda _{0}}{2\pi }\sum_{p=\pm
1}pD_{m}^{(0p)}H_{m}(\lambda _{0}\rho ),  \label{magnetic0b}
\end{eqnarray}%
with the coefficients
\end{subequations}
\begin{equation}
D_{m}^{(0p)}=\frac{\pi }{2ic}\left[ v_{\perp }J_{m+p}(\lambda _{0}\rho
_{0})-v_{\parallel }\frac{\lambda _{0}}{k_{z}}J_{m}(\lambda _{0}\rho _{0})%
\right] .  \label{Dm0p}
\end{equation}%
The corresponding expressions for $\rho <\rho _{0}$ are obtained from (\ref%
{magnetic0}) by the replacements $J\rightleftarrows H$ of the Bessel and
Hankel functions. The part $H_{ml}^{(1)}$ in (\ref{Hdecompose}) is due to
the inhomogeneity and is given by formulae
\begin{subequations}
\label{magnetic}
\begin{eqnarray}
H_{ml}^{(1)} &=&-\frac{qk_{z}}{2\pi i^{\sigma _{l}}}\sum_{p=\pm 1}p^{\sigma
_{l}-1}D_{m}^{(p)}J_{m+p}(\lambda _{0}\rho ),\;l=\rho ,\phi ,
\label{magnetica} \\
H_{mz}^{(1)} &=&-\frac{q\lambda _{0}}{2\pi }\sum_{p=\pm
1}pD_{m}^{(p)}J_{m}(\lambda _{0}\rho ),  \label{magneticb}
\end{eqnarray}%
where the notation
\end{subequations}
\begin{equation}
D_{m}^{(p)}=C_{m}^{(p)}-\frac{i\pi v_{\parallel }\lambda _{0}}{2ck_{z}}\frac{%
W_{m}^{H}}{W_{m}^{J}}J_{m}(\lambda _{0}\rho _{0}),\quad p=\pm 1,  \label{Dm}
\end{equation}%
is introduced. By making use of the Maxwell equation $\nabla \times \mathbf{H%
}=-i\omega \varepsilon _{0}\mathbf{E}/c$, one can derive the corresponding
Fourier coefficients for the electric field:
\begin{subequations}
\label{electric}
\begin{eqnarray}
E_{ml}^{(1)} &=&\frac{qci^{1-\sigma _{l}}}{4\pi \omega
_{m}(k_{z})\varepsilon _{0}}\sum_{p=\pm 1}p^{\sigma _{l}}J_{m+p}(\lambda
_{0}\rho )\left[ \left( \frac{\omega _{m}^{2}(k_{z})\varepsilon _{0}}{c^{2}}%
+k_{z}^{2}\right) D_{m}^{(p)}-\lambda _{0}^{2}D_{m}^{(-p)}\right]
,
\label{electrica} \\
E_{mz}^{(1)} &=&\frac{qic\lambda _{0}k_{z}}{2\pi \omega
_{m}(k_{z})\varepsilon _{0}}\sum_{p=\pm 1}D_{m}^{(p)}J_{m}(\lambda _{0}\rho
),  \label{electricb}
\end{eqnarray}%
where $l=\rho ,\phi $. As it follows from these formulae, $\mathbf{E}%
_{m}^{(1)}\cdot \mathbf{H}_{m}^{(1)}=0$, i.e., the corresponding Fourier
components of the electric and magnetic fields are perpendicular to each
other.

Taking in the formulae given above $\omega _{0}=0$ for a fixed $\rho _{0}$,
as a special case we obtain the fields generated by a charge moving with a
constant velocity $v_{\parallel }$ on a straight line $\rho =\rho _{0}$
parallel to the cylinder axis. For this case one has
\end{subequations}
\begin{equation}
\lambda _{i}^{2}=\lambda _{i}^{(0)2}\equiv k_{z}^{2}(\beta
_{i\parallel }^{2}-1),\;\beta _{i\parallel }=v_{\parallel
}\sqrt{\varepsilon _{i}}/c,\;i=0,1,  \label{lambdai2m0}
\end{equation}%
and $v_{\perp }=0$. As a result the dependence on the parameter $\rho _{0}$
in the coefficients $D_{m}^{(p)}$ is in the form of the Bessel function $%
J_{m}(\lambda _{0}\rho _{0})$. It follows from here that in the limit $\rho
_{0}\rightarrow 0$ (the particle moves along the axis of the dielectric
cylinder) the term with $m=0$ contributes only. The latter property is a
simple consequence of the azimuthal symmetry of the corresponding problem.

The formulae for the fields are simplified for the $m=0$ mode. In this case $%
\lambda _{i}=\lambda _{i}^{(0)}$ and for the function $\alpha _{m}$ from (%
\ref{bet1}) one has%
\begin{equation}
\alpha _{0}=\frac{\varepsilon _{1}\lambda _{0}J_{0}(\lambda _{0}\rho
_{1})H_{0}^{\prime }(\lambda _{1}\rho _{1})-\varepsilon _{0}\lambda
_{1}J_{0}^{\prime }(\lambda _{0}\rho _{1})H_{0}(\lambda _{1}\rho _{1})}{%
(\varepsilon _{1}-\varepsilon _{0})W_{1}^{J}}.  \label{alfa0}
\end{equation}%
The expression for the coefficients $D_{0}^{(p)}$ takes the form%
\begin{equation}
D_{0}^{(p)}=p\frac{i\pi v_{\perp }W_{1}^{H}}{2cW_{1}^{J}}J_{1}(\lambda
_{0}\rho _{0})-\frac{v_{\parallel }J_{0}(\lambda _{0}\rho _{0})}{cW_{0}^{J}}%
\left[ \frac{i\pi \lambda _{0}}{2k_{z}}W_{0}^{H}+\frac{k_{z}}{\rho _{1}}%
\frac{H_{0}(\lambda _{1}\rho _{1})H_{0}^{\prime }(\lambda _{1}\rho _{1})}{%
\alpha _{0}W_{1}^{J}}\right] .  \label{D0p}
\end{equation}%
For the corresponding Fourier components of the electric field we obtain the
formulae%
\begin{equation}
E_{0l}^{(1)}=\frac{qck_{z}\beta _{0\parallel }^{2(\sigma _{l}-1)}}{2\pi
i^{\sigma _{l}-1}v_{\parallel }\varepsilon _{0}}J_{1}(\lambda _{0}\rho
)\sum_{p=\pm 1}p^{\sigma _{l}-1}D_{0}^{(p)},\;E_{0z}^{(1)}=\frac{iqc\lambda
_{0}^{(0)}k_{z}}{2\pi v_{\parallel }\varepsilon _{0}}J_{0}(\lambda _{0}\rho
)\sum_{p=\pm 1}D_{0}^{(p)},  \label{E0l1}
\end{equation}%
with $l=\rho ,\phi $ and $\sigma _{l}$ is defined after formulae (\ref%
{GFcomp}). These formulae are used in the next section to the derive the
formula for the corresponding radiation intensity inside the cylinder.

\section{Radiation fields inside a dielectric cylinder}

\label{sec:radfields}

In this section we consider the radiation field propagating inside a
dielectric cylinder. First of all let us show that the part corresponding to
the fields of the charge in a homogeneous medium with permittivity $%
\varepsilon _{0}$ does not contribute to the radiation field in this region.
This directly follows from the estimate of the integral over $k_{z}$ in the
corresponding formula (\ref{vecpot2}) on the base of the stationary phase
method. As in the integral over $k_{z}$ the phase $k_{z}z$ has no stationary
points and the integrand is a function of the class $C^{\infty }(R)$, for
large values $|z|$ the integral vanishes more rapidly than any power of $%
1/|z|$. From this argument it follows that the radiation field is determined
by the singular points of the integrand in the integral over $k_{z}$. In the
expressions for the parts of the Fourier components $H_{ml}^{(1)}$ and $%
E_{ml}^{(1)}$, the coefficients $D_{m}^{(p)}$ enter in the form of
combinations $\sum_{p}D_{m}^{(p)}$ and $\sum_{p}pD_{m}^{(p)}$. By using
formulae (\ref{Dm}), it can be seen that for $m\neq 0$ these combinations
are regular at the points corresponding to the zeros of the functions $%
W_{m}^{J}$ and $W_{m\pm 1}^{J}$. The only poles of the parts of the Fourier
components for the fields determined by relations (\ref{magnetic}), (\ref%
{electric}) are the zeros of the function $\alpha _{m}$ appearing in the
denominators of Eqs. (\ref{Bm+p}), (\ref{Cm+p}). It can be seen that this
function has zeros only for $\lambda _{1}^{2}<0$. Note that for the
corresponding modes in the exterior region (see Ref. \cite{Saha05}), $\rho
>\rho _{1}$, the Fourier coefficients are proportional to the MacDonald
function $K_{\nu }(|\lambda _{1}|\rho )$, $\nu =m,m\pm 1$, and they are
exponentially damped in the region outside the cylinder. These modes are the
eigenmodes of the dielectric cylinder and propagate inside the cylinder. By
using the properties of the cylindrical functions, for $m\neq 0$ the
function $\alpha _{m}$ can also be written in the form%
\begin{equation}
\alpha _{m}=\frac{U_{m}}{(\varepsilon _{1}-\varepsilon _{0})\left(
V_{m}^{2}-m^{2}u^{2}\right) },  \label{alfmnew}
\end{equation}%
where we have used the notations%
\begin{eqnarray}
V_{m} &=&|\lambda _{1}|\rho _{1}\frac{J_{m}^{\prime }}{J_{m}}+\lambda
_{0}\rho _{1}\frac{K_{m}^{\prime }}{K_{m}},\;u=\lambda _{0}/|\lambda
_{1}|+|\lambda _{1}|/\lambda _{0}  \label{Vmrad} \\
U_{m} &=&V_{m}\left( \varepsilon _{0}|\lambda _{1}|\rho _{1}\frac{%
J_{m}^{\prime }}{J_{m}}+\varepsilon _{1}\lambda _{0}\rho _{1}\frac{%
K_{m}^{\prime }}{K_{m}}\right) -m^{2}\frac{\lambda _{0}^{2}+|\lambda
_{1}|^{2}}{\lambda _{0}^{2}|\lambda _{1}|^{2}}\left( \varepsilon _{1}\lambda
_{0}^{2}+\varepsilon _{0}|\lambda _{1}|^{2}\right) .  \label{Umrad}
\end{eqnarray}%
Here and below it is understood $K_{m}=K_{m}(|\lambda _{1}|\rho _{1})$, $%
J_{m}=J_{m}(\lambda _{0}\rho _{1})$ if the argument of the function is
omitted, and the prime means the differentiation with respect to the
argument of the function. Now the equation for the eigenmodes is written in
the standard form (see, for instance,~\cite{Jackson})%
\begin{equation}
U_{m}=0.  \label{eigmodesnew2}
\end{equation}%
Unlike to the case of the waveguide with perfectly conducting walls, here
the eigenmodes with $m\neq 0$\ are not decomposed into independent TE and TM
parts. This decomposition takes place only for the $m=0$ mode and this case
will be discussed below separately.

We denote by $\lambda _{0}\rho _{1}=\lambda _{m,s}$, $s=1,2,\ldots $, the
solutions to equation (\ref{eigmodesnew2}) with respect to $\lambda _{0}\rho
_{1}$ for the modes with $m\neq 0$. The corresponding modes $%
k_{z}=k_{m,s}^{(\pm )}$ are related to these solutions by the formula%
\begin{equation}
k_{m,s}^{(\pm )}=\frac{m\omega _{0}\sqrt{\varepsilon _{0}}}{c\left( 1-\beta
_{0\parallel }^{2}\right) }\left[ \beta _{0\parallel }\pm \sqrt{%
1+b_{m,s}^{2}\left( \beta _{0\parallel }^{2}-1\right) }\right] ,\;b_{m,s}=%
\frac{c\lambda _{m,s}}{m\omega _{0}\rho _{1}\sqrt{\varepsilon _{0}}}.
\label{kmspm}
\end{equation}%
These modes are real under the condition $b_{m,s}^{2}\left( 1-\beta
_{0\parallel }^{2}\right) \leqslant 1$. For $\beta _{0\parallel }<1$, the
condition for $k_{m,s}^{(\pm )}$ to be real defines the maximum value for $s$%
, which we will denote by $s_{m}$:%
\begin{equation}
\lambda _{m,s_{m}}<\frac{m\omega _{0}\rho _{1}\sqrt{\varepsilon _{0}}}{c%
\sqrt{1-\beta _{0\parallel }^{2}}}<\lambda _{m,s_{m}+1}.  \label{smdef}
\end{equation}%
If the Cherenkov condition $\beta _{0\parallel }>1$ is satisfied, the upper
limit $s_{m}$ for $s$ is determined by the dispersion law for the dielectric
permittivity via the condition $\varepsilon _{0}(\omega
_{m})>c^{2}/v_{\parallel }^{2}$. The function $\omega _{m}(k_{z})$ is also
expressed in terms of $\lambda _{m,s}$:%
\begin{equation}
\omega _{m}(k_{z})=\frac{m\omega _{0}}{1-\beta _{0\parallel }^{2}}\left[
1\pm \beta _{0\parallel }\sqrt{1+b_{m,s}^{2}\left( \beta _{0\parallel
}^{2}-1\right) }\right] .  \label{omegams}
\end{equation}

In order to obtain unambiguous result for the fields we should specify the
integration contour in the complex plane $k_{z}$ in Eq. (\ref{vecpot2}). For
this we note that in physically real situations the dielectric permittivity $%
\varepsilon _{0}$ is a complex quantity: $\varepsilon _{0}=\varepsilon
_{0}^{\prime }+i\varepsilon _{0}^{\prime \prime }$. Assuming that $%
\varepsilon _{0}^{\prime \prime }$ is small, this induces an imaginary part
for $k_{z}$ given by formula%
\begin{equation}
{\mathrm{Im\,}}k_{z}=\pm \frac{\left[ 1\pm \beta _{0\parallel }\sqrt{%
1+b_{m,s}^{2}(\beta _{0\parallel }^{2}-1)}\right] ^{2}}{2\sqrt{\varepsilon
_{0}}(1-\beta _{0\parallel }^{2})^{2}\sqrt{1+b_{m,s}^{2}(\beta _{0\parallel
}^{2}-1)}}i\varepsilon _{0}^{\prime \prime }(\omega _{m}),  \label{Imkz}
\end{equation}%
where the coefficient and the argument of $\varepsilon _{0}^{\prime \prime
}(\omega _{m})$ are evaluated with the real part of dielectric permittivity.
Note that one has $\varepsilon _{0}^{\prime \prime }(\omega _{m})\gtrless 0$
for $\omega _{m}\gtrless 0$. In the case $\beta _{0\parallel }<1$ from (\ref%
{omegams}) we have $\omega _{m}>0$ and, hence, in accordance with (\ref{Imkz}%
), ${\mathrm{Im\,}}k_{z}\gtrless 0$ for the modes $k_{m,s}^{(\pm )}$.
Deforming the integration contour we see that for $\beta _{0\parallel }<1$
in the integral over $k_{z}$ in (\ref{vecpot21}), the contour avoids the
poles $k_{m,s}^{(-)}$ from above and the poles $k_{m,s}^{(+)}$ from below.
For the case $\beta _{0\parallel }>1$ we have $\omega _{m}<0$ for the modes
with $k_{z}=k_{m,s}^{(+)}$ and $\omega _{m}>0$ for the modes with $%
k_{z}=k_{m,s}^{(-)}$ and, hence, in accordance with formula (\ref{Imkz}) for
both types of modes ${\mathrm{Im\,}}k_{z}<0$. As a result, in this case in
the integral over $k_{z}$ all poles $k_{m,s}^{(\pm )}$ should be avoided
from above.

Now let us consider the electromagnetic fields inside the cylinder for large
distances from the charge. In this case the main contribution into the
fields comes from the poles of the integrand. Having specified the
integration contour now we can evaluate the radiation parts of the fields
inside the cylinder. Closing the integration contour by the large
semicircle, under the condition $\beta _{0\parallel }<1$ one finds%
\begin{equation}
F_{l}^{\mathrm{(rad)}}(\mathbf{r},t)=\sigma F_{l}^{(\sigma )}(\mathbf{r}%
,t)=\sigma 4\pi {\mathrm{Re}}\left[ i\sum_{m=1}^{\infty }e^{im(\phi -\omega
_{0}t)}\sum_{s=1}^{s_{m}}\underset{k_{z}=k_{m,s}^{(\sigma )}}{\mathrm{Res}}%
e^{ik_{z}(z-v_{\parallel }t)}F_{ml}(k_{z},\rho )\right] ,  \label{Alpoles}
\end{equation}%
where $\sigma =+(-)$ for $z-v_{\parallel }t>0$ ($z-v_{\parallel }t<0$). This
expression describes waves propagating along the positive direction of the
axis $z$ for $\sigma =+$ and for $\sigma =-$, $b_{m,s}<1$, and waves
propagating along the negative direction to the axis $z$ for $\sigma =-$, $%
1<b_{m,s}<1/\sqrt{1-\beta _{0\parallel }^{2}}$. Under the condition $\beta
_{0\parallel }>1$ for the radiation fields one finds%
\begin{equation}
F_{l}^{\mathrm{(rad)}}(\mathbf{r},t)=-\sum_{\sigma =\pm }F_{l}^{(\sigma )}(%
\mathbf{r},t)\theta (v_{\parallel }t-z),  \label{Flrad2}
\end{equation}%
where $\theta (x)$ is the Heaviside unit step function. As we could expect,
in this case the radiation field is behind the charge. Formula (\ref{Flrad2}%
) describes the waves propagating along the positive direction of the axis $%
z $ for $\sigma =+$ and for $\sigma =-$, $b_{m,s}>1$, and waves propagating
along the negative direction of the axis $z$ for $\sigma =-$, $b_{m,s}<1$.
Note that for $b_{m,s}>1$ there are no waves propagating along the negative
direction of the axis $z$.

By evaluating the residue in (\ref{Alpoles}), for the functions $%
F_{l}^{(\sigma )}(\mathbf{r},t)$ one finds
\begin{equation}
F_{l}^{(\sigma )}(\mathbf{r},t)=4\pi {\mathrm{Re}}\left[ i\sum_{m=1}^{\infty
}e^{im(\phi -\omega _{0}t)}\sum_{s=1}^{s_{m}}e^{ik_{z}(z-v_{\parallel }t)}%
\frac{F_{ml}^{(\sigma )}(k_{z},\rho )}{d\alpha _{m}/dk_{z}}\right]
_{k_{z}=k_{m,s}^{(\sigma )}},  \label{Alalpha}
\end{equation}%
where the formulae for the coefficients $F_{ml}^{(\sigma )}(k_{z},\rho )$, $%
l=\rho ,\phi ,z$ are obtained from the corresponding expressions for $%
F_{ml}^{(\sigma )}$ by the replacement $C_{m}^{(p)}\rightarrow
C_{m}^{(p\sigma )}$ for the components of the vector potential and by the
replacement $D_{m}^{(p)}\rightarrow C_{m}^{(p\sigma )}$ for the electric and
magnetic fields, with%
\begin{equation}
C_{m}^{(p\sigma )}=\frac{K_{m+p}K_{m}}{4pcW_{m+p}^{(\sigma )}}\left[
4v_{\parallel }k_{m,s}^{(\sigma )}\rho _{1}\frac{J_{m}(\lambda _{m,s}\rho
_{01})}{W_{m}^{(\sigma )}}-v_{\perp }\lambda _{m,s}^{(\sigma )}\sum_{l=\pm 1}%
\frac{J_{m+l}(\lambda _{m,s}\rho _{01})}{lW_{m+l}^{(\sigma )}}\right] ,
\label{Cmpalpha}
\end{equation}%
where we have introduced the notation%
\begin{equation}
W_{m}^{(\sigma )}=\lambda _{m,s}^{(\sigma )}J_{m}(\lambda
_{m,s})K_{m}^{\prime }(\lambda _{m,s}^{(\sigma )})-\lambda
_{m,s}J_{m}^{\prime }(\lambda _{m,s})K_{m}(\lambda _{m,s}^{(\sigma )}),
\label{Wmsig}
\end{equation}%
and%
\begin{equation}
\lambda _{m,s}^{(\pm )}=\sqrt{\left( 1-\frac{\varepsilon _{1}}{\varepsilon
_{0}}\right) k_{m,s}^{(\pm )2}\rho _{1}^{2}-\frac{\varepsilon _{1}}{%
\varepsilon _{0}}\lambda _{m,s}^{2}},\;\rho _{01}=\frac{\rho _{0}}{\rho _{1}}%
.  \label{lambmsplmin}
\end{equation}%
Note that we have the following relations%
\begin{equation}
W_{m+p}^{(\sigma )}=\left( pV_{m}-mu\right) J_{m}K_{m},\;W_{m}^{(\sigma )}=%
\frac{k_{m,s}^{(\sigma )2}\rho _{1}^{2}}{\lambda _{m,s}\lambda
_{m,s}^{(\sigma )}V_{m}}\left( V_{m}^{2}-m^{2}u^{2}\right) J_{m}K_{m}.
\label{WsigtoV}
\end{equation}%
Unfortunately, for the case $m\neq 0$ we have no summation formula over the
eigenmodes like to that for the $m=0$ mode (see below) and for the further
evaluation of the radiation fields numerical methods have to be used.

Now we turn to the modes with $m=0$. In this case there are two types of
eigenmodes. The first ones are the zeros of the function $\alpha _{0}$
corresponding to the zeros of the nominator in (\ref{alfa0}). The equation
for these eigenmodes is written in the form
\begin{equation}
\varepsilon _{0}|\lambda _{1}|\frac{J_{0}^{\prime }}{J_{0}}+\varepsilon
_{1}\lambda _{0}\frac{K_{0}^{\prime }}{K_{0}}=0,\;\lambda _{i}=\lambda
_{i}^{(0)},  \label{eigmodesm0}
\end{equation}%
where $\lambda _{i}^{(0)}$ is defined by formula (\ref{lambdai2m0}). For the
presence of these modes we have the conditions $\beta _{1\parallel }<1<\beta
_{0\parallel }$. In particular we should have $\varepsilon _{1}<\varepsilon
_{0}$. From formula (\ref{D0p}) it follows that the combination $%
\sum_{p}pD_{m}^{(p)}$ has no poles at these eigenmodes and for the
corresponding radiation field one has $H_{0\rho }^{(1)}=H_{0z}^{(1)}=0$ and $%
E_{0\phi }^{(1)}=0$. As a result, these waves are TM waves or the waves of
the E-type. The values for $k_{z}$ defined by equation (\ref{eigmodesm0}) we
will denote by $\pm k_{0,s}^{\mathrm{(E)}}$, $s=1,2,\ldots $. The second
type of $m=0$ eigenmodes corresponds to the zeros of the function $W_{1}^{J}$%
. The corresponding equation can be written in the form
\begin{equation}
\lambda _{0}\frac{K_{0}^{\prime }}{K_{0}}+|\lambda _{1}|\frac{J_{0}^{\prime }%
}{J_{0}}=0,\;\lambda _{i}=\lambda _{i}^{(0)}.  \label{eigmodesm0M}
\end{equation}%
These zeros are poles for the combination $\sum_{p}pD_{m}^{(p)}$, whereas
the combination $\sum_{p}D_{m}^{(p)}$ is regular at these zeros. For the
corresponding radiation fields one has $H_{0\phi }^{(1)}=0$ and $E_{0\rho
}^{(1)}=E_{0z}^{(1)}=0$ and these waves are TE waves or the waves of the
M-type. The corresponding eigenvalues for $k_{z}$ we will denote by $\pm
k_{0,s}^{\mathrm{(M)}}$, $s=1,2,\ldots $. Note that both types of the
eigenmodes are also obtained from (\ref{eigmodesnew2}) taking $m=0$.

As in the case of $m\neq 0$, to define the fields from formula (\ref%
{vecpot21}) we should specify the contour for the integration in the complex
plane $k_{z}$. In the way similar to that used for the modes with $m\neq 0$,
it can be seen that the contour avoids the poles $\pm k_{0,s}^{\mathrm{(F)}}$
from above. The corresponding radiation fields are found by evaluating the $%
k_{z}$-integrals with the help of residue theorem. For $z-v_{\parallel }t>0$
we close the integration contour in the upper half-plane of the complex
variable $k_{z}$ and the radiation fields are zero. We could expect this
result as $\beta _{0\parallel }>1$ and the radiation field is behind the
charge. For the points in the region $z-v_{\parallel }t<0$ we close the
integration contour in the lower half-plane and the integral is equal to the
residues at the poles $\pm k_{0,s}^{\mathrm{(E)}}$ and $\pm k_{0,s}^{\mathrm{%
(M)}}$ \ multiplied by $-2\pi i$. Here we will present the formulae for the
nonzero components of the electric field:%
\begin{eqnarray}
E_{\rho ,m=0}^{\mathrm{(E)}} &=&\frac{4q}{\rho _{1}^{2}}\sum_{s}\frac{%
\varepsilon _{1}\sqrt{\beta _{0\parallel }^{2}-1}}{\varepsilon
_{1}-\varepsilon _{0}}\frac{J_{0}(\rho _{0}\lambda _{0,s}^{\mathrm{(E)}%
})J_{1}(\rho \lambda _{0,s}^{\mathrm{(E)}})\sin \left[ k_{0,s}^{\mathrm{(E)}%
}(z-v_{\parallel }t)\right] }{\varepsilon _{0}J_{0}^{\prime 2}(\rho
_{1}\lambda _{0,s}^{\mathrm{(E)}})+\varepsilon _{1}(\beta _{0\parallel
}^{2}-1)J_{0}^{2}(\rho _{1}\lambda _{0,s}^{\mathrm{(E)}})},  \notag \\
E_{z,m=0}^{\mathrm{(E)}} &=&\frac{4q}{\rho _{1}^{2}}\sum_{s}\frac{%
\varepsilon _{1}(\beta _{0\parallel }^{2}-1)}{\varepsilon _{1}-\varepsilon
_{0}}\frac{J_{0}(\rho _{0}\lambda _{0,s}^{\mathrm{(E)}})J_{0}(\rho \lambda
_{0,s}^{\mathrm{(E)}})\cos \left[ k_{0,s}^{\mathrm{(E)}}(z-v_{\parallel }t)%
\right] }{\varepsilon _{0}J_{0}^{\prime 2}(\rho _{1}\lambda _{0,s}^{\mathrm{%
(E)}})+\varepsilon _{1}(\beta _{0\parallel }^{2}-1)J_{0}^{2}(\rho
_{1}\lambda _{0,s}^{\mathrm{(E)}})},  \label{m0waves} \\
E_{\phi ,m=0}^{\mathrm{(M)}} &=&\frac{4qv_{\perp }}{\rho
_{1}^{2}v_{\parallel }}\sum_{s}\frac{J_{1}(\rho _{0}\lambda _{0,s}^{\mathrm{%
(M)}})J_{1}(\rho \lambda _{0,s}^{\mathrm{(M)}})}{(\varepsilon
_{1}-\varepsilon _{0})J_{0}^{\prime 2}(\rho _{1}\lambda _{0,s}^{\mathrm{(M)}%
})}\cos \left[ k_{0,s}^{\mathrm{(M)}}(z-v_{\parallel }t)\right] ,  \notag
\end{eqnarray}%
where we have introduced the notation%
\begin{equation}
\lambda _{0,s}^{\mathrm{(F)}}=k_{0,s}^{\mathrm{(F)}}\sqrt{\beta _{0\parallel
}^{2}-1},\;\mathrm{F=E,M}.  \label{lamb0F}
\end{equation}%
In these formulae the summands with a given $s$ describe the radiation field
with the frequency $\omega _{0,s}^{\mathrm{(F)}}=v_{\parallel }k_{0,s}^{%
\mathrm{(F)}}$. In Appendix it is shown that $\lambda _{0,s}^{\mathrm{(F)}%
}\rho _{1}\gtrsim 1$ and, hence, for the corresponding frequencies one has
the estimate $\omega _{0,s}^{\mathrm{(F)}}\gtrsim v_{\parallel }/(\rho _{1}%
\sqrt{\beta _{0\parallel }^{2}-1})$. In formulae (\ref{m0waves}) the upper
limit for the summation over $s$, which we will denote by $s_{m}$, is
determined by the dispersion law of the dielectric permittivity $\varepsilon
=\varepsilon (\omega )$ through the condition $\varepsilon (\omega _{0,s}^{%
\mathrm{(F)}})>c^{2}/v_{\parallel }^{2}$.

Having the electric field we can evaluate the radiation intensity on the
mode $m=0$ inside the dielectric waveguide by using the formula%
\begin{equation}
I_{m=0}=-\int \left( j_{\phi }E_{\phi ,m=0}+j_{z}E_{z,m=0}\right) \rho d\rho
d\varphi dz.  \label{Im0}
\end{equation}%
For this quantity one finds%
\begin{equation}
I_{m=0}=I_{m=0}^{\mathrm{(E)}}+I_{m=0}^{\mathrm{(M)}},  \label{Im=0}
\end{equation}%
where for the contributions of the TM and TM waves we have%
\begin{subequations}
\label{I}
\begin{eqnarray}
I_{m=0}^{\mathrm{(E)}} &=&\frac{2q^{2}v_{\parallel }}{\rho _{1}^{2}}\sum_{s}%
\frac{\beta _{0\parallel }^{2}-1}{\varepsilon _{0}-\varepsilon _{1}}\frac{%
J_{0}^{2}(\rho _{0}\lambda _{0,s}^{\mathrm{(E)}})}{(\varepsilon
_{0}/\varepsilon _{1})J_{0}^{\prime 2}(\rho _{1}\lambda _{0,s}^{\mathrm{(E)}%
})+(\beta _{0\parallel }^{2}-1)J_{0}^{2}(\rho _{1}\lambda _{0,s}^{\mathrm{(E)%
}})},  \label{IE} \\
I_{m=0}^{\mathrm{(M)}} &=&\frac{2q^{2}v_{\perp }^{2}}{\rho
_{1}^{2}v_{\parallel }}\sum_{s}\frac{J_{1}^{2}(\rho _{0}\lambda _{0,s}^{%
\mathrm{(M)}})}{(\varepsilon _{0}-\varepsilon _{1})J_{0}^{\prime 2}(\rho
_{1}\lambda _{0,s}^{\mathrm{(M)}})}.  \label{IM}
\end{eqnarray}
\end{subequations}
As the modes $k_{0,s}^{\mathrm{(F)}}$ are not explicitly known as functions
on $s$, formulae (\ref{IE}), (\ref{IM}) are not convenient for the
evaluation of the corresponding radiation intensities. More convenient form
may be obtained by making use of the summation formula (\ref{sumformdiel})
derived in Appendix, taking
\begin{equation}
\eta =\sqrt{\frac{1-\beta _{1\parallel }^{2}}{\beta _{0\parallel }^{2}-1}}.
\label{eta}
\end{equation}%
In formula (\ref{sumformdiel}) we choose $\alpha =\varepsilon
_{0}/\varepsilon _{1}$, $f(z)=zJ_{0}^{2}(z\rho _{0}/\rho _{1})$ for the
waves of the E-type and $\alpha =1$, $f(z)=zJ_{1}^{2}(z\rho _{0}/\rho _{1})$
for the waves of the M-type. For both types of modes one has $f(-ix)=-f(ix)$
and we can use the version of the summation formula given by (\ref%
{sumformdiel1}). In the intermediate step of the calculations, it
is technically simpler instead of considering the dispersion of
the dielectric permittivity to assume that in formulae (\ref{I}) a
cutoff function $\psi _{\mu }(\lambda _{0,s}^{\mathrm{(F)}})$ is
introduced with $\mu $ being the cutoff parameter and $\psi
_{0}=1$ [for example, $\psi _{\mu }(x)=\exp (-\mu x)$], which will
be removed after the summation. In this way, after the application
of the summation formula we find the following
results%
\begin{subequations}
\label{I2}
\begin{eqnarray}
I_{m=0}^{\mathrm{(E)}} &=&q^{2}v_{\parallel }\left[ c^{2}\int_{0}^{\infty }dx%
\frac{x}{\varepsilon _{0}}\psi _{\mu }(x)J_{0}^{2}(\rho _{0}x)+\frac{4}{\pi
^{2}\rho _{1}^{2}}\int_{0}^{\infty }dx\,\frac{I_{0}^{2}(x\rho _{0}/\rho _{1})%
}{\varepsilon _{1}xg_{\varepsilon _{0}/\varepsilon _{1}}(\eta ,x)}\right] ,
\label{IE2} \\
I_{m=0}^{\mathrm{(M)}} &=&\frac{q^{2}v_{\perp }^{2}v_{\parallel }}{c^{2}}%
\left[ \int_{0}^{\infty }dx\frac{x\psi _{\mu }(x)}{\beta _{0\parallel }^{2}-1%
}J_{1}^{2}(\rho _{0}x)-\frac{4}{\pi ^{2}\rho _{1}^{2}}\int_{0}^{\infty }dx\,%
\frac{I_{1}^{2}(x\rho _{0}/\rho _{1})}{(\beta _{0\parallel
}^{2}-1)xg_{1}(\eta ,x)}\right] ,  \label{IM2}
\end{eqnarray}%
\end{subequations}
where the function $g_{\alpha }(\eta ,x)$ is defined by formula (\ref%
{denomdiel}). In the first terms in the square brackets replacing the
integration variable by the frequency $\omega =v_{\parallel }x/\sqrt{\beta
_{0\parallel }^{2}-1}$ and introducing the physical cutoff \ through the
condition $\beta _{0\parallel }>1$ instead of the cutoff function, we see
that these terms coincide with the radiation intensities on the harmonic $m=0
$ for the waves of the E- and M-type in the homogeneous medium with
dielectric permittivity $\varepsilon _{0}$ (see Ref. \cite{Saha05}). The
second terms in the square brackets are induced by the inhomogeneity of the
medium in the problem under consideration. Note that, unlike to the terms
corresponding to the homogeneous medium, for $\rho _{0}<\rho _{1}$ the
latter are finite also in the case when the dispersion is absent: for large
values of $x$ the integrands decay as $\exp [-2x(1-\rho _{0}/\rho _{1})]$
(for this reason we have removed the cutoff function from these terms). In
particular, from the last observation it follows that under the condition $%
(1-\rho _{0}/\rho _{1})\gg v_{\parallel }/\omega _{d}$, where
$\omega _{d}$ is the characteristic frequency for the dispersion
of the dielectric permittivity, the influence of the dispersion on
the inhomogeneity induced terms can be neglected. Note that in a
homogeneous medium the corresponding radiation propagates under
the Cherenkov angle $\theta _{C}=\arccos (1/\beta _{0\parallel })$
and has a continuous spectrum, whereas the radiation described by
(\ref{I2}) propagates inside the dielectric
cylinder and has a discrete spectrum with frequencies $\omega _{0,s}^{%
\mathrm{(F)}}$. As the function $g_{\alpha }(\eta ,x)$ is always
nonnegative, from formulae (\ref{I2}) we conclude that the
presence of the cylinder amplifies the $m=0$ part of the radiation
for the waves of the E-type and suppresses the the radiation for
the waves of the M-type. In the helical undulators one has
$v_{\perp }\ll v_{\parallel }$ and the contribution of the TM
waves dominates.

\begin{figure}[tbph]
\begin{center}
\begin{tabular}{cc}
\epsfig{figure=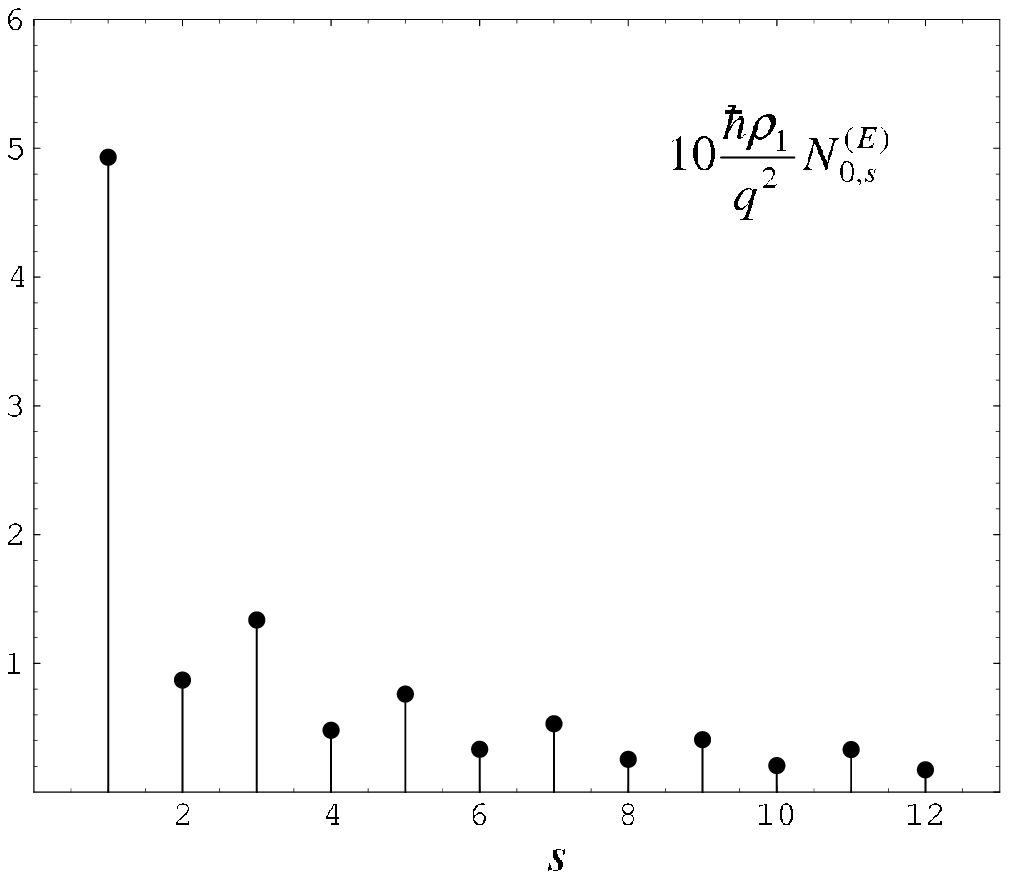,width=6.5cm,height=6cm} & \quad %
\epsfig{figure=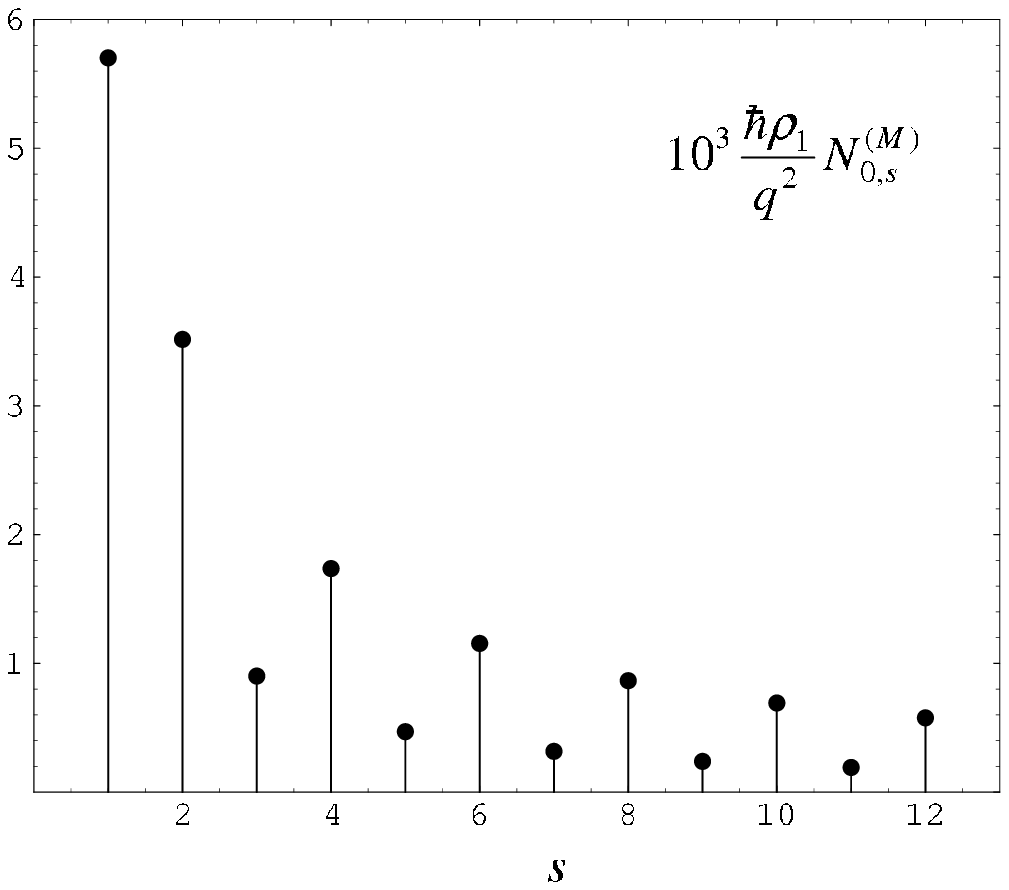,width=6.5cm,height=6cm}%
\end{tabular}%
\end{center}
\caption{The number of the radiated quanta on the mode $m=0$ as a
function of the harmonic number $s$ for the waves of the E- (left
panel) and M-type (right panel) (for the values of the parameters
see the text).} \label{fig1}
\end{figure}

In addition to the total intensity, it is of interest to have the
distribution of the intensity over the frequencies. To this aim we define
the partial intensities $I_{m=0,s}^{\mathrm{(F)}}$ for a given $s$ related
to the quantity $I_{m=0}^{\mathrm{(F)}}$ by the formula $I_{m=0}^{\mathrm{(F)%
}}=\sum_{s}I_{m=0,s}^{\mathrm{(F)}}$. In figure \ref{fig1} we have presented
the results of the numerical evaluation for the number of the quanta
radiated per unit time, $N_{m=0,s}^{\mathrm{(F)}}=I_{m=0,s}^{\mathrm{(E)}%
}/\hbar \omega _{0,s}^{\mathrm{(F)}}$, on a harmonic with a given $s$, as a
function of $s$ for the electron with the energy 2~MeV moving inside a
cylinder with dielectric permittivity $\varepsilon _{0}=3$. It is assumed
that for the exterior region $\varepsilon _{1}=1$. The angle $\theta _{0}$
between the electron velocity and the cylinder axis ($\tan \theta
_{0}=v_{\perp }/v_{\parallel }$) is taken 0.1 rad, and $\rho _{0}/\rho
_{1}=0.5$.
The values of the parameter $\rho _{1}\lambda _{0,s}^{\mathrm{%
(F)}}$ for the first eight harmonics are presented in the table. The
corresponding frequencies are related to this parameter by the formula $%
\omega _{0,s}^{\mathrm{(F)}}=v_{\parallel }\lambda _{0,s}^{\mathrm{(E)}}/%
\sqrt{\beta _{0\parallel }^{2}-1}$.

\begin{center}
\begin{tabular}{|c|c|c|c|c|c|c|c|c|}
\hline
$s$ & 1 & 2 & 3 & 4 & 5 & 6 & 7 & 8 \\ \hline
$\rho _{1}\lambda _{0,s}^{\mathrm{(E)}}$ & 2.728 & 5.931 & 9.102 & 12.262 &
15.415 & 18.565 & 21.713 & 24.860 \\ \hline
$\rho _{1}\lambda _{0,s}^{\mathrm{(M)}}$ & 2.507 & 5.653 & 8.801 & 11.947 &
15.092 & 18.236 & 21.380 & 24.523 \\ \hline
\end{tabular}
\end{center}

\section{Conclusion}

\label{sec:Conc}

We have considered the electromagnetic field generated by a charged particle
moving along a helical orbit inside a dielectric cylinder immersed into a
homogeneous medium. The fields and the spectral-angular distribution of the
radiation intensity in the exterior medium have been investigated in our
previous paper \cite{Saha05}. In particular, the conditions were specified
under which strong narrow peaks appear in the angular distribution for the
number of radiated quanta. In the present paper we study the fields inside
the cylinder. By using the corresponding formulae for the components of the
Green function, we have derived expressions for the electromagnetic
potentials and fields. These expressions are presented in the form of sums
of two terms. The first ones correspond to the fields generated by the
charge in a homogeneous medium with the same dielectric permittivity as that
for the material of the cylinder. The second terms are induced by the
difference of the dielectric permittivities for the exterior and interior
media, and are given by formulae (\ref{vecpot3}), (\ref{magnetic}), (\ref%
{electric}). We have extracted the parts of the fields which are
responsible for the radiation. For the radiation propagating
inside the cylinder the projection of the wave vector on the
cylinder axis takes discrete set of values which are solutions to
the eigenmode equation (\ref{eigmodesnew2}). The expressions for
the electromagnetic fields corresponding to the radiation
propagating inside the cylinder are derived. As an application we
have considered the radiation intensity on the mode $m=0$. In this
case the radiation field is separated into purely TE and TM modes
and we have found the intensities for both types of these modes
determined by formulae (\ref{I}). These formulae presents the
intensities as sums over the modes of the dielectric waveguide.
However, as the corresponding eigenmodes are not explicitly known,
this form is not convenient for the numerical evaluation of the
radiation intensity. In Appendix, by using the generalized
Abel-Plana formula, we derive a summation formula for the series
over the normal modes of the dielectric cylinder which allows to
extract from the radiation intensity the parts corresponding to
the radiation in a homogeneous medium and to present the
inhomogeneity-induced part in terms of exponentially converging
integrals. Unlike to the parts corresponding to the homogeneous
medium, the terms induced by the presence of the cylinder are
finite also in the case when the dispersion is absent. We have
specified the condition under which the influence of dispersion on
the inhomogeneity-induced effects can be neglected.

\section*{Acknowledgement}

The authors are grateful to Professor A. R. Mkrtchyan for general
encouragement and to Professor L. Sh. Grigoryan, S. R. Arzumanyan, H. F.
Khachatryan for stimulating discussions. The work has been supported by
Grant No.~0063 from Ministry of Education and Science of the Republic of
Armenia and in part by PVE/CAPES Program.

\appendix

\section{Summation formula over the eigenmodes}

In this section we derive a summation formula for the series over zeros of
the function%
\begin{equation}
C_{\alpha }(\eta ,z)=V_{\alpha }\{J_{0}(z),K_{0}(\eta z)\},  \label{Cdiel}
\end{equation}%
where and in what follows for given functions $F(z)$ and $G(z)$ we use the
notation%
\begin{equation}
V\{F(z),G(\eta z)\}=F(z)G^{\prime }(\eta z)+\alpha \eta F^{\prime }(z)G(\eta
z),  \label{Vfgnotdiel}
\end{equation}%
with $\alpha \geqslant 1$ and $\eta $ being real constants. For
this we will use the generalized Abel-Plana formula from
\cite{Saha00} (for applications of the generalized Abel-Plana
formula in quantum field theory with boundaries see
\cite{Saha06}). In this formula
we substitute%
\begin{equation}
g(z)=if(z)\frac{V_{\alpha }\left\{ Y_{0}(z),K_{0}(\eta z)\right\} }{%
C_{\alpha }(\eta ,z)}.  \label{gdiel}
\end{equation}%
For the combinations of the functions entering in the generalized Abel-Plana
formula one has%
\begin{equation}
f(z)-(-1)^{k}g(z)=f(z)\frac{V_{\alpha }\{H_{0}^{(k)}(z),K_{0}(\eta z)\}}{%
C_{\alpha }(\eta ,z)},  \label{fmingdiel}
\end{equation}%
where $H_{\nu }^{(k)}(z)$, $k=1,2$, are the Hankel functions. Let us denote
positive zeros of the function $C_{\alpha }(\eta ,z)$ by $k_{s}$, $%
s=1,2,\ldots $, assuming that these zeros are arranged in the ascending
order. Note that, for $z\gg 1$ we have $C_{\alpha }(\eta ,z)\approx
K_{0}^{\prime }(\eta z)-\alpha \eta zK_{0}(\eta z)/2<0$ and, hence, $%
k_{s}\gtrsim 1$. By using the asymptotic formulae for the cylindrical
functions for large values of the argument, it can be seen that for large
values $s$ one has $k_{s}\approx -\arctan (1/\alpha \eta )+\pi /4+\pi s$.
For the derivative of the function $C_{\alpha }(\eta ,z)$ at the zeros $k_{s}
$\ one obtains%
\begin{equation}
C_{\alpha }^{\prime }(\eta ,z)=-\eta \frac{K_{0}(\eta z)}{J_{0}(z)}\left[
\alpha (1+\alpha \eta ^{2})J_{0}^{\prime 2}(z)+(\alpha -1)J_{0}^{2}(z)\right]
,\;z=k_{s}.  \label{Cderdiel}
\end{equation}%
In particular, it follows from here that the zeros are simple. Assuming that
the function $f(z)$ is analytic in the right half-plane, for the residue
term in the generalized Abel-Plana formula one finds%
\begin{equation}
\underset{z=k_{s}}{\mathrm{Res}}g(z)=-\frac{i}{\pi }T_{\alpha
}(k_{s})f(k_{s}),  \label{Resgdiel}
\end{equation}%
where we have introduced the notation%
\begin{equation}
T_{\alpha }(z)=\frac{2\alpha /z}{\alpha (1+\alpha \eta ^{2})J_{0}^{\prime
2}(z)+(\alpha -1)J_{0}^{2}(z)}.  \label{Talf}
\end{equation}%
Substituting the expressions for the separate terms into the generalized
Abel-Plana formula we obtain the following result%
\begin{eqnarray}
\lim_{x_{0}\rightarrow \infty }\left[ \sum_{s=1}^{s_{0}}T_{\alpha
}(k_{s})f(k_{s})-\int_{0}^{x_{0}}dx\,f(x)\right]  &=&-\frac{1}{\pi }%
\int_{0}^{\infty }dx\,\left[ f(ix)\frac{V_{\alpha
}\{K_{0}(x),H_{0}^{(2)}(\eta x)\}}{V_{\alpha }\{I_{0}(x),H_{0}^{(2)}(\eta
x)\}}\right.   \notag \\
&&\left. +f(-ix)\frac{V_{\alpha }\{K_{0}(x),H_{0}^{(1)}(\eta x)\}}{V_{\alpha
}\{I_{0}(x),H_{0}^{(1)}(\eta x)\}}\right] ,  \label{sumformdiel}
\end{eqnarray}%
where $s_{0}$ is defined by the relation $k_{s_{0}}<x_{0}<k_{s_{0}+1}$. This
formula is valid for functions $f(z)$ obeying the condition
\begin{equation}
|f(z)|<\epsilon (x)e^{c|y|},\;z=x+iy,\;|z|\rightarrow \infty ,
\label{condfz}
\end{equation}%
where $c<2$ and $\epsilon (x)\rightarrow 0$ for $x\rightarrow \infty $.
Formula (\ref{sumformdiel}) is further simplified for functions satisfying
the additional condition $f(-ix)=-f(ix)$:%
\begin{equation}
\lim_{x_{0}\rightarrow \infty }\left[ \sum_{s=1}^{s_{0}}T_{\alpha
}(k_{s})f(k_{s})-\int_{0}^{x_{0}}dx\,f(x)\right] =-\frac{4i\alpha }{\pi ^{2}}%
\int_{0}^{\infty }dx\,\frac{f(ix)}{x^{2}g_{\alpha }(\eta ,x)},
\label{sumformdiel1}
\end{equation}%
where%
\begin{eqnarray}
g_{\alpha }(\eta ,x) &=&I_{0}^{2}(x)\left[ J_{1}^{2}(\eta x)+Y_{1}^{2}(\eta
x)\right] +\alpha ^{2}\eta ^{2}I_{1}^{2}(x)\left[ J_{0}^{2}(\eta
x)+Y_{0}^{2}(\eta x)\right]   \notag \\
&&-2\alpha \eta I_{0}(x)I_{1}(x)\left[ J_{0}(\eta x)J_{1}(\eta x)+Y_{0}(\eta
x)Y_{1}(\eta x)\right] .  \label{denomdiel}
\end{eqnarray}%
Note that we have denoted $g_{\alpha }(\eta
,x)=|V\{I_{0}(z),H_{0}^{(1)}(\eta z)\}|^{2}$ and this function is always
nonnegative.

\end{document}